\documentclass[sn-mathphys,Numbered,pdflatex]{sn-jnl}% Math and Physical Sciences Reference Style
\usepackage{graphicx}%
\usepackage{multirow}%
\usepackage{amsmath,amssymb,amsfonts}%
\usepackage{amsthm}%
\usepackage{mathrsfs}%
\usepackage[title]{appendix}%
\usepackage{xcolor}%
\usepackage{textcomp}%
\usepackage{manyfoot}%
\usepackage{lmodern}%
\usepackage{oplotsymbl}
\usepackage{booktabs}%
\usepackage{algorithm}%
\usepackage{algorithmicx}%
\usepackage{algpseudocode}%
\usepackage{listings}%
\newtheorem{theorem}{Theorem}%  meant for continuous numbers
\newtheorem{proposition}[theorem]{Proposition}% 

\begin{document}

\title[Article Title]{\centering Negation and Identity in a Modal Mode Theory \\or\\the Persistence of the Negative\footnote{An invocation of \cite{noys2010persistence}}}


\author{\fnm{Juan} \sur{Afanador}}\email{jafanad@gmail.com}

% \affil*[1]{\orgdiv{Department}, \orgname{Organization}, \orgaddress{\street{Street}, \city{City}, \postcode{100190}, \state{State}, \country{Country}}}




% \date{31 August 2023}

\maketitle

\section*{}\label{sec1}
\vspace{-1cm}
This piece threads substructurality and modality into a negation that activates the downside of equivalence and identity in a fibrational framework \cite{licata2017fibrational, de2016fibrational, gratzer2020multimodal}. The piece is a working through of negation and contradiction as type-theoretic/categorial objects, towards an immanent critique of the subtending univalent paradigm (v. \cite{awodey2018univalence}). Although this is not the terminus of the piece, i wish to try and delineate the epistemic and intra-mundane problematics intertwined therewith. 
% -- sociation warrants that thinking subverts itself. 

The piece's terminus is a mode theory of an intuitionistic modal logic $\mathcal{L}_{nod}$ that internalises a restriction on the Double Negation Elimination (DNE) rule. The meta-negations of $\mathcal{L}_{nod}$ give way to a strictly weak $\star-autonomous$ \cite{pastro2010note, cockett2006coherence} (sc. Grothendieck-Verdier \cite{boyarchenko2013duality}) category, which warrants a comonadic characterisation of contradiction. The proposed comonadic semantics may be applied to model theory \cite{abramsky2017pebbling}, deductive complexity \cite{abramsky2021comonadic}, or combinatorics \cite{abramsky2022comonadic}, but it is in displaying  the tensions between negation (sc. the negative) and identity (sc. the affirmative), inherent in such instrumentalisms, where the relevance of the semantics may lie. 

To work my case, i use an intuitionistic modal language akin to $LBiKt$ under standard notation and Kripke semantics \cite{postniece2010proof}. That is, $\mathcal{M}=\langle W, \smallfrown, \smallsmile, \mathcal{I} \rangle$ is a model defined by the frame formed by a set of worlds $W$, world-accessibility relations $\smallfrown, \smallsmile\subseteq W\times W$ \cite{onishi2015substructural, lahav2017sequent}, and an interpretative mapping $\mathcal{I}:Atom(\mathcal{L}_{nod}) \mapsto 2^W$, whereupon satisfiability and validity are as habitually posed. $\mathcal{L}_{nod}$ extends $\mathcal{L}_{BiN}$ (v. \cite{ciabattoni2014hypersequent} and \cite{odintsov2021routley}) to accommodate the operators $\{\blacktriangleright, \blacktriangleleft, \triangleright, \triangleleft\}$, allowing for a modal (tense) split of negation -- e.g., $\mathcal{M}, w\vDash\blacktriangleright A,\, \exists v(v\smallsmile w\, \text{and}\, \mathcal{M},v\nvDash A)$ and $\mathcal{M}, w\vDash\triangleright A,\, \forall v(v\smallfrown w\, \text{then}\, \mathcal{M},v\nvDash A)$ for $w, v\in W$. (In)exhaustiveness as (the) failure (of the concept to model its object) is key.

Contexts in $\mathcal{L}_{nod}$ are of two types, either $X::=I|\alpha\in \mathcal{L}_{nod}| \sharp X|\flat X|X\pentagodot Y|X>Y$ or $\Gamma[\,]::=[\,]|\Gamma[\,]\pentagodot (X)| (X)\pentagodot \Gamma[\,]$. Following the $LBiInt_1$ usage, $X > \Gamma[\,]$ is a negative context. Structural and object-language operators can be mapped, with little overload, like so: $I\mapsto \{\top, \bot\}$, $\pentagodot\,\mapsto \{\land, \lor\}$, $>\,\mapsto \{-\hspace{-.8em}<, \to\}$, $\sharp\mapsto\{\blacktriangleright, \triangleright\}$, and $\flat\mapsto\{\blacktriangleleft, \triangleleft\}$. Exemplar rules of the attendant sequent calculus are $x:\flat A \vdash Y/x:\blacklozenge A \vdash Y$ and $\Gamma, F_{x\pentagodot y}(x:\blacklozenge A, y:\flat A)/\Gamma, x:\blacklozenge A$.

In our mode theory, frames index modes, $\{\sharp, \flat\}$ connote their modalities, and maps between them prefigure type-theoretic 2-cells. Alternatively, these are the objects, morphisms, and (categorical) 2-cells of the $\mathbf{Struct}$ 2-category \cite{abramsky2022comonadic}. In the forthcoming piece i lift the category these modes inhabit to a monadic (display) category (viz. \cite{ahrens2019displayed}), internalising the notion of contexts noted above to show that DNE must needs not be. Or less informally:

\begin{proposition} \label{prop}
There exists, at least, one mode $A\in \mathbf{Struct}$, such that the canonical morphism of the symmetric closed monoidal category $\langle \mathbf{Struct}, \pentagodot, I, \to \rangle$ is not an isomorphism.
\end{proposition}

To prove Proposition \ref{prop}, the dualising unit of the monoidal category is derived from the modal split of negation and the contextual interpretation of modalities. Although algebraically germane, the dualising techniques of Priestley \cite{gehrke2014canonical} and Esakia \cite{celani2014easkia} inform our derivation of duality. Once this result is established, the covariant functor described by the dualising unit serves to probe for modes which cannot be identified with their double negations. Or more eloquently, we are interested in the fragments of the nameless mode theory that Proposition \ref{prop} inhabits, where what is negated stays negated.

In closing i rehearse the opening caveat -- the subversion of (adjunctor) equivalence, in our mode-theoretic rendition of negation, also constitutes a point of entry into a questioning of the Univalence Axiom as a principle of logic \cite{awodey2018univalence}. In unfolding this piece two themes are obliquely interweaved: the metonymic claim of essential sameness that the univalent paradigm articulates in the computarised idiom of a post-theoretical world; and the adjacent abolition of the transcendental subject from logic (sc. cognition), whereby the intervening agency between individual action and social norm is subl(im)ated. Thinking negativity (sc. negation) may rehabilitate a critical engagement with the (social) aporias of what is. 

\bibliography{neging}% common bib file

\end{document}